\documentclass[pra,twocolumn,showpacs,a4paper]{revtex4}
\usepackage{graphicx}\usepackage{amsmath,amssymb}
\usepackage{hyperref}
\def\SU#1{\mathbb{SU}(#1)}
\def\d{\operatorname{d}}\def\<{\langle}\def\>{\rangle}\def\Tr{\operatorname{Tr}}
\def\kk{\rangle\!\rangle}\def\bb{\langle\!\langle}
\begin{document}
\title{Optimal phase estimation for qubit mixed states}

\author{Giacomo Mauro D'Ariano}\email{dariano@unipv.it} \author{Chiara
  Macchiavello}\email{chiara@unipv.it} \affiliation{{\em QUIT Group}
  of the INFM, unit\`a di Pavia} \homepage{http://www.qubit.it}
\affiliation{Dipartimento di Fisica ``A. Volta'', via Bassi 6, I-27100
  Pavia, Italy}\author{Paolo
  Perinotti}\email{perinotti@fisicavolta.unipv.it} \affiliation{{\em
    QUIT Group} of the INFM, unit\`a di Pavia}
\homepage{http://www.qubit.it} \affiliation{Dipartimento di Fisica
  ``A. Volta'', via Bassi 6, I-27100 Pavia, Italy} \date{\today}

\begin{abstract}
We present the optimal phase estimation for qubits in mixed states, for an arbitrary number of
qubits prepared in the same state. 
\end{abstract}
 \maketitle
\section{Introduction}
The problem of measuring the quantum phase has been a very long standing one in Quantum Mechanics,
since London's\cite{london} and Dirac's first attempts \cite{dirac} in the late twenties. One of the
main motivations is that the estimation of the phase shift experienced by a quantum system is the
only way of measuring time with high precision in Quantum Mechanics, since we lack a time
observable. This posed the problem of quantum phase estimation naturally within the framework of
frequency standards based on atomic clocks \cite{atclocks}, and more generally, in high precision
measurements and interferometry, the typical scenario in which the sensitivity of phase estimation
is profitably used. 

More recently, the encoding of information into the relative phase of quantum systems is exploited in quantum 
computation and communication. In fact, in quantum computing most of the existing quantum algorithms
can be regarded as multiparticle interferometers, with the output of the  
computation encoded in the relative phase between different paths \cite{cemm}. On the other hand, in
some cryptographic communication protocols (e. g. BB84\cite{BB84}) information is encoded into
phase properties. 

\par The above numerous applications had focused a great deal of interest on the problem of optimal
phase estimation, which has been widely studied in a thousand of papers (see for example
Ref. \cite{pst})  since the beginning of quantum theory \cite{london,dirac}. The first
satisfactory partial solution of the problem appeared 
in the late 70's (see Refs. \cite{helstrom} and \cite{holevo} for reviews), and these works are generally 
regarded as one of the major successes of quantum estimation theory and covariant measurements,
allowing a first consistent definition of phase, without the problems suffered by the original definition
proposed by Dirac \cite{dirac} in terms of an alleged observable conjugated to the number operator of the
harmonic oscillator. 

In the covariant treatment of Ref. \cite{holevo} the estimated parameter is a phase shift resulting
from the action of a circle group of unitary transformations, with generator a selfadjoint
operator with purely integer spectrum. A generalization of this method to degenerate phase-shift
generator has been presented in Ref. \cite{dms}. Such general approach can be applied to any input
pure state, along with a restricted class of mixed states, the so called {\em phase-pure states}
\cite{dms,tesi}. 

The possibility of efficiently estimating the phase for mixed states is of fundamental interest for practical 
implementations, in the presence of unavoidable noise which generally turns pure states into
mixed, and for estimation of local phase-shift on entangled states. As a matter of fact, the freedom
in the choice of the optimal measurement which results from degenerate shift operators\cite{dms}
opens the problem of the stability of the quality of the estimation with increasing mixing of the shifted state. 
The problem of optimal phase estimation on mixed states is also very relevant conceptually, the phase being 
one of the most elusive quantum concepts. The main reason why the problem  of optimal phase estimation
on mixed states has never been addressed systematically so far is due to the intrinsic technical difficulties 
faced in any quantum estimation problem with mixed states. In this paper we derive the optimal measurement
for phase estimation on qubits in mixed states, for an arbitrary number $N$ of qubits prepared in
the same state, using either the Uhlman fidelity or the periodicized variance as a figure of merit.
\section{Theoretical derivation}
Let us consider a system of $N$ identical qubits prepared in the same
mixed state $\rho_{\vec n}=\frac12(I+\vec n\cdot\vec\sigma)$, where
$|\vec n|\doteq r<1$ and $\sigma_i$ are the three Pauli matrices. The
total state of the $N$ qubits is described by the density matrix
$R_{\vec n}=\rho_{\vec n}^{\otimes N}$. The phase transformation
$U_\phi$ is generated by the $z$ component of the total angular momentum
$J_z=\frac12\sum_{k=1}^N\sigma_z^{(k)}$, namely
\begin{equation}
R_{\vec n}(\phi)=U_\phi R_{\vec n}U_\phi^\dag=\left[e^{-i\frac\phi2\sigma_z} \rho_{\vec n}
  e^{i\frac\phi2\sigma_z}\right]^{\otimes N}\label{shift}
\end{equation}
The problem is now to estimate the unknown phase-shift $\phi$ on the known state $R_{\vec n}$.
We consider a covariant measurement, namely we require that the efficiency of the measurement
procedure does not depend on the value of the phase to be estimated. In this case, without loss of
generality we can assume that the initial state $\rho_{\vec n}$ has no component along $\sigma_y$,
corresponding to real matrix $\rho_{\vec n}$ in the $\sigma_z$ representation. The phase estimation
problem then resorts to find the best POVM \cite{helstrom}  
$P(\d\phi)$ for determining the unknown parameter $\phi$
in Eq. (\ref{shift}). The fact that $P(\d\phi)$ is a POVM corresponds to the constraints
\begin{equation}
P(\d\phi)\geq0\,,\quad\int_0^{2\pi}P(\d\phi)=I.\label{eq:norpovm}
\end{equation}
In the quantum estimation approach the optimality is defined by maximizing the average of a given figure of
merit $C(\phi,\phi^\prime)$, assuming a uniform prior distribution of the parameter $\phi$
\begin{equation}
\<C\>=\int_0^{2\pi}\frac{\d\phi}{2\pi}\int_0^{2\pi}C(\phi,\phi^\prime)\Tr[U_\phi R_{\vec n}U^\dag_\phi P(\d\phi^\prime)]\,,
\end{equation}
where $C(\phi,\phi^\prime)=C(\phi-\phi^\prime)$. In Ref. \cite{holevo} it was proved that the
solution for an estimation problem covariant under a unitary group representation can be written as
the group orbit under the same representation of a fixed positive operator $\xi$ (called {\em seed} of the
POVM), and for the present case one has 
\begin{equation}
P(\d\phi^\prime)=U_{\phi^\prime}\xi
U_{\phi^\prime}^\dag\frac{\d\phi^\prime}{2\pi}\,.
\end{equation}
In the following we will denote by $|m,a\>$ an orthonormal basis, with $m$ denoting the eigenvalues
of $\frac12 J_z$, which label the equivalence classes of irreducible representations of the
group $\{U_\phi\}$, while $a$ is a degeneration index, corresponding to the multiplicity space of the representation $m$. 
The normalization condition \eqref{eq:norpovm} for the POVM $P(\d\phi)$ implies that 
$\<m,a|\xi|m,b\>=0$ for $a\neq b$, whereas $\<m,a|\xi|m,a\>=1$ for all $a$.
\par In quantum estimation theory the quantity to be minimized can be always written in the form of
the expectation of a cost operator as follows
\begin{equation}
\<C\>=\int_0^{2\pi}\frac{\d\phi^\prime}{2\pi} C(\phi^\prime)\Tr[R_{\vec n} U_{\phi^\prime}\xi U^\dag_{\phi^\prime}]\,.
\end{equation}
The choice of the cost function $C(\phi)$ depends on the estimation criterion. The  most commonly
adopted criteria are the periodicized variance 
\begin{equation}
C(\phi)\equiv v(\phi)=4\sin^2\frac\phi 2=2(1-\cos\phi),
\end{equation}
or the (opposite of the) fidelity between the true and the estimated states, which for mixed states
has the well known Uhlman's form\cite{Uhlman}
\begin{equation}
C(\phi)\equiv 1-F(\phi)=1-\left[\Tr\sqrt{\sqrt{U_\phi\rho_{\vec n} U^\dag_\phi}
\rho_{\vec n}\sqrt{U_\phi\rho_{\vec n}
    U^\dag_\phi}}\right]^2\,,\label{eq:Uhlman}
\end{equation}
which for qubits simplifies as follows\cite{hubner}
\begin{equation}
1-F(\phi)=\frac{r^2}{2}(1-\cos\phi).
\end{equation}
Both cost functions depend on $\phi$ only through its cosine, whence we need to maximize the
averaged $\cos\phi$, namely 
\begin{equation}
\<c\>=\frac12\int_0^{2\pi}\frac{\d\phi^\prime}{2\pi} (e^{i\phi^\prime}+e^{-i\phi^\prime})\Tr[R_{\vec n} U_{\phi^\prime}\xi U^\dag_{\phi^\prime}]\,.
\label{avc}
\end{equation}
The evaluation of the integral in Eq. (\ref{avc}) leads to the following expression
\begin{equation}
\<c\>=\mbox{Re}\sum_{m,a,b}\<m,a|\xi|m+1,b\>\<m+1,b|R_{\vec n}|m,a\>\,.
\label{eq:avc}
\end{equation}
We now decompose $\rho^{\otimes N}$ into irreducible representations of $\SU 2$, as shown in Ref. 
\cite{cimaek} recasting $R_{\vec n}$ into invariant block-diagonal form on the orthonormal basis
$|j,m,\alpha\>_{\vec  b}=U_{j,\alpha}|j,m,1\>_{\vec b}$ for the minimal invariant subspaces of the
$\SU 2$ representations, with $\vec b=\frac{\vec n}r$, and $U_{j,\alpha}$ denoting a suitable set of
unitary operators.
\begin{align}
  &R_{\vec n}\doteq\rho_{\vec n}^{\otimes N}=\sum_{j=\bb N/2\kk}^J(r_+r_-)^J
\sum_{\alpha=1}^{d_j}U_{j,\alpha}\tau_{j,1} U_{j,\alpha}^\dag\,,\label{eq:rhon}\\
  &\tau_{j,1}=\sum_{m=-j}^j\left(\frac{r_+}{r_-}\right)^m|j,m,1\>_{\vec b}\<j,m,1|\,,\label{eq:blocks}\\
  &|j,m,1\>_{\vec b}=|j,m\>_{\vec b}\otimes|\Psi_-\>^{\otimes J-j}\,,\label{singlets}
\end{align}
where $r_\pm\doteq \frac{1}{2}(1\pm r)$,
$\bb x\kk$ denotes the fractional part of $x$ (i. e. $\bb N/2\kk=0$ for $N$ even and $\bb
N/2\kk=1/2$ for $N$ odd), $J=N/2$, and $d_j$ is the multiplicity of the $j$-th irreducible
representation of $\SU 2$ 
\begin{equation}
d_j=\binom{2J}{J-j}-\binom{2J}{J-j-1}\,,
\end{equation}
whereas $|\Psi_-\>$ denotes the singlet state. This decomposition is
useful since $m,j,\alpha$ label also the irreducible representations
of $\{U_\phi\}$, $m$ being the eigenvalue of $J_z$, and $j,\alpha$ becoming both
degeneration indices. The block diagonal form of $R_{\vec n}$ shows that the only coupling produced
by the phase shift between irreducible components with $m$ and $m+1$ can occur only between vectors
in the same invariant subspace $j,\alpha$ of $\SU2$. Upon recasting $R_{\vec n}$ in the form of Eq. 
\eqref{eq:rhon}, the value of $\<c\>$ in Eq. \eqref{eq:avc} involves
only the following terms
\begin{equation}
\<c\>=\mbox{Re}\sum_{m,j\alpha}\<m,j\alpha|\xi|m+1,j\alpha\>\<m+1,j\alpha|R_{\vec n}|m,j\alpha\>\,,
\label{eq:bdavc}
\end{equation}
where we used the short notation $|m,j\alpha\>\doteq|j,m,\alpha\>_z$, since the subspaces $j,\alpha$
are invariant under any unitary in $\SU 2$, and $|j,m,\alpha\>_{\vec
  b}=T^{(j)}(g)|j,m,\alpha\>_z$ for some $g\in\SU 2$.
Now, the following bounding hold
\begin{eqnarray}
\<c\>&\leq&\left|\sum_{m,j\alpha}\<m,j,\alpha|\xi|m+1,j\alpha\>\<m+1,j\alpha|R_{\vec n}|m,j\alpha\>\right|\nonumber\\
&\leq&\sum_{m,j\alpha}\left|\<m,j,\alpha|\xi|m+1,j\alpha\>\<m+1,j\alpha|R_{\vec n}|m,j\alpha\>\right|\nonumber\\
&\leq&\sum_{m,j\alpha}\left|\<m+1,j\alpha|R_{\vec n}|m,j\alpha\>\right|\,,
\label{eq:bounds}
\end{eqnarray}
where the last bound follows from positivity of $\xi$. We show now that all bounds can be achieved by a
suitable choice of the operator $\xi$ compatible with constraints (\ref{eq:norpovm}). The first
two bounds can indeed be 
achieved by choosing the phases of the matrix elements
$\<m,j,\alpha|\xi|m+1,j\alpha\>$ in such a way that they compensate
the corresponding phases of $\<m+1,j\alpha|R_{\vec n}|m,j\alpha\>$.
The last bound is achieved by just taking the moduli of the matrix
elements $\<m,j\alpha|\xi|m+1,j\alpha\>$ to be 1. It remains to prove
that these choices are compatible with positivity. In order to show
this, let us write
\begin{equation}
\<m+1,j\alpha|R_{\vec n}|m,j\alpha\>=|\<m+1,j\alpha|R_{\vec
  n}|m,j\alpha\>|e^{i\chi(m+1,m,j\alpha)}\,. 
\end{equation}
Since only the elements on the first over-diagonal and under-diagonal
are involved, one can write the phases $\chi(m+1,m,j\alpha)$ as the difference of two functions as follows
\begin{equation}
\chi(m+1,m,j\alpha)=\gamma(m,j\alpha)-\gamma(m+1,j\alpha)\,,
\label{eq:linph}
\end{equation}
as the number of independent linear equations in Eq. \eqref{eq:linph}
is $2^N-1$ while the unknown phases are $2^N$.  Then one can take
\begin{equation}
\xi=\sum_{j,\alpha}|e(j,\alpha)\>\<e(j,\alpha)|\,, 
\label{eq:optpovm}
\end{equation}
where $|e(j,\alpha)\>$ is the generalized Susskind-Glogower vector
\begin{equation}
|e(j,\alpha)\>=\sum_{m=-j}^je^{i\gamma(m,j\alpha)}|m,j,\alpha\>\,.\label{Susskind}
\end{equation}
It is immediate that Eq. \eqref{eq:optpovm} represents a positive operator
and by construction $\xi$ produces a normalized POVM, while achieving the bounding in
Eq. \eqref{eq:bounds}
\par Specifically, for a collection of identically prepared mixed initial states, we have 
\begin{eqnarray}
\<c\>&=&\sum_{m,j,\alpha}\left|\<m+1,j\alpha|R_{\vec n}|m,j\alpha\>\right|\nonumber\\
&=&\sum_{m,j,\alpha}(r_+r_-)^J\times\nonumber\\
&&\left|\sum_n\left(\frac{r_+}{r_-}\right)^n\<j,m+1,\alpha|j,n,\alpha\>_{\vec b}\<j,n,\alpha|j,m,\alpha\>\right|\nonumber\\
&=&\sum_{j=\bb N/2\kk}^J\sum_{m=-j}^jd_j(r_+r_-)^J\times\nonumber\\ &&
\left|\sum_n\left(\frac{r_+}{r_-}\right)^nT^{(j)}(g_{\vec b}
)_{m+1,n}T^{(j)}(g_{\vec b})^\dag_{n,m}\right|\,.
\label{eq:avcos}
\end{eqnarray}
Notice that since we assumed that $\vec n$ has no component along the
direction $y$, then $g_{\vec b}$ is just the rotation around the axis $y$
connecting the oriented $z$ axis with $\vec b$, namely $T(g_{\vec b})=e^{i\theta J_y}$
for some $\theta$, with $J_y=\frac12\sum_{k=1}^N\sigma_y^{(k)}$.

\section{Numerical results}
The expression for the Wigner matrix elements $T^{(j)}(g_{\vec b})_{lk}$
is given by \cite{messiah}
\begin{equation}
\begin{split}
T^{(j)}(g_{\vec b})_{lk}=&\sum_t(-1)^t\frac{\sqrt{(j+l)!(j-l)!(j+k)!(j-k)!}}{(j+l-t)!(j-k-t)!(t-l+k)!t!}\times\\
&\cos^{2j+l-k}\frac\theta2\sin^{2t-l+k}\frac\theta2\,.
\end{split}
\end{equation}
\begin{figure}[h]
\includegraphics[width=1.6in]{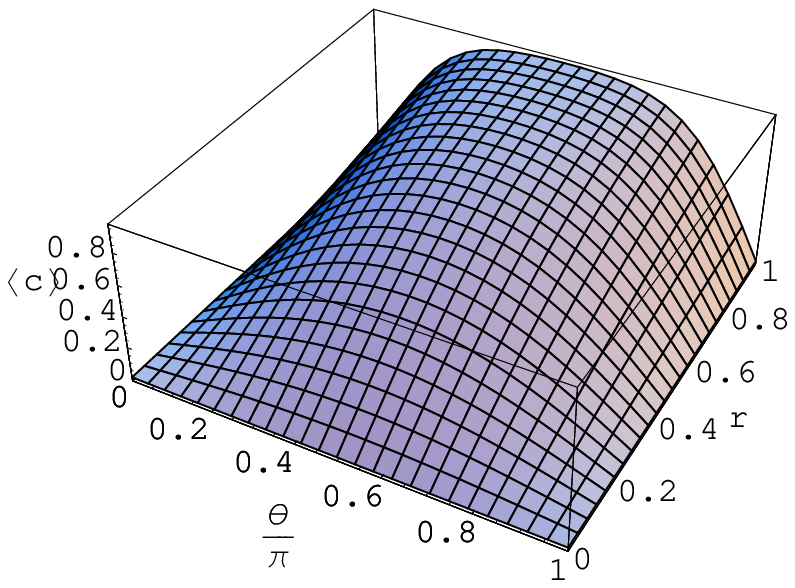}\hfill\includegraphics[width=1.6in]{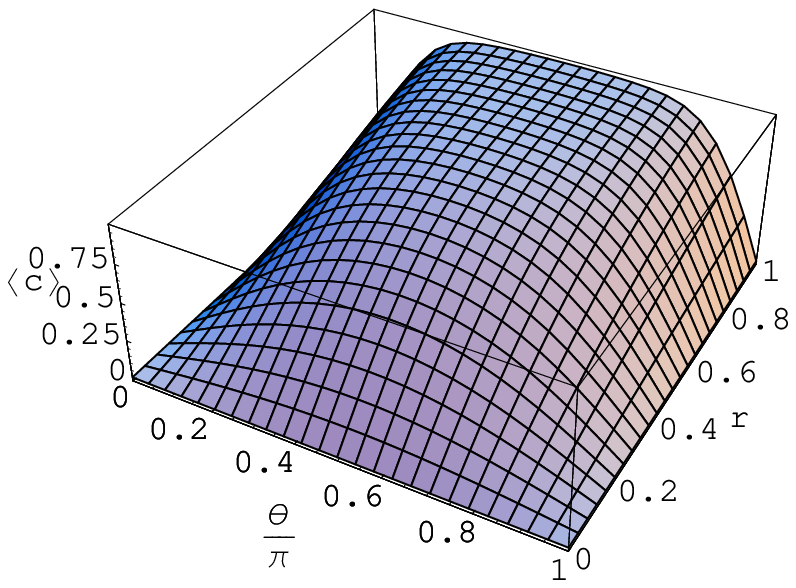}
\caption{The plot of $\<c\>$ as a function of $\theta$ and $r$. The plots correspond to systems of 10 qubits and 20 qubits respectively.\label{f:plot}}
\end{figure}
The explicit expression of Eq. \eqref{eq:avcos} is very lengthy, and
has been evaluated using symbolic calculus for $J$ up to 21/2, namely
for a total number of spins equal to 21. The plot of the averaged
cosine $\<c\>$ as a function of $\theta$ and $r$ is represented in
Fig. \ref{f:plot} and exhibits two interesting intuitive features. The
first is that the maximum versus $\theta$ occurs for
$\theta=\frac\pi2$, namely for qubits lying in the equatorial plane.
The second is the improving figure of merit versus the purity $r$.
Equatorial pure qubits are optimal for phase detection, however, the
figure of merit is quite stable around its maxima, still with $N=10$
copies.\par
\begin{figure}[h]
  \includegraphics[width=2.5in]{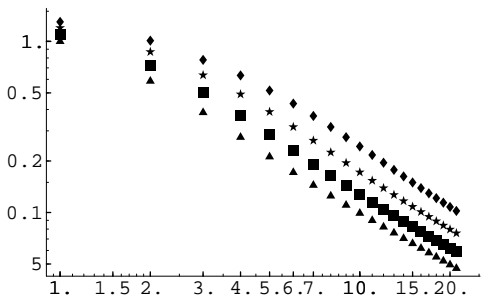}
\caption{The logarithmic plots represent $2(1-\<c\>)$, where $\<c\>$ is the averaged cosine, as a function of the number of spins $N$, for $\theta=\frac\pi2$ and for the following values of $r$: $\blacklozenge\ r=.7$, $\bigstar\ r=.8$, $\blacksquare\ r=.9 $, $\blacktriangle\ r=1$.\label{f:gra}}
\end{figure}
Fig. \ref{f:gra} shows the averaged cosine $\<c\>$ versus the
number of qubits $N$ for equatorial states. Numerically, for
$N\to\infty$ we find the asymptotic behavior $2(1-\<c\>)\propto
N^{-1}$. More precisely, for the Uhlman fidelity $F$ in Eq.
(\ref{eq:Uhlman}) we find an asymptotic behavior saturating the
Cramer-Rao lower bound \cite{cramer}. This gives a strict lower bound
for variance $\Delta\phi^2$ valid for any estimate. For independent
copies, one has \cite{helstrom}
\begin{equation}
\Delta\phi^2 \ge \frac{1}{N} \Tr[(\partial\rho/\partial\phi)\mathcal L]^{-1}
\end{equation}
where for each $\phi$ the operator $\mathcal L$ is defined by the identity 
\begin{equation}
\partial\rho/\partial\phi\doteq\frac12(\rho\mathcal L+\mathcal L\rho)\,.
\end{equation}
Notice that the bound holds for {\em any} estimate, whence regardless the nature of the measurement
(corresponding to either joint or separable POVM's). Since the estimation is covariant, we can just
consider $\phi=0$.  
A simple evaluation shows that $\mathcal L=r\cos\theta\sigma_y$, and the bound is then given by
$\frac 1N\Tr[\rho_0\mathcal L^2]^{-1}=\frac 1{Nr^2\cos^2\theta}$, namely
\begin{equation}
\Delta\phi^2 \ge \frac{1}{N} \frac 1{r^2\cos^2\theta}.
\end{equation}
\begin{figure}[h]
  \includegraphics[width=2.5in]{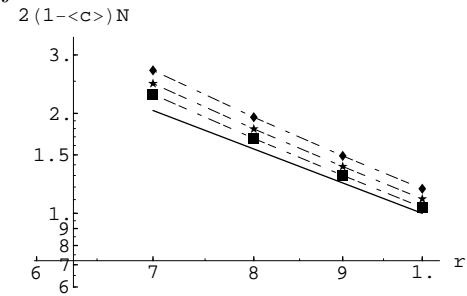}
\caption{The logarithmic plot of $2N(1-\<c\>)$ vs $r$, for $N=16,18,20$ and $\theta=0$. The line on the bottom represents the bound given by the Cramer-Rao inequality, namely $1/r^2$}
\label{f:asym}
\end{figure}
For small $\Delta\phi^2 $ using the Taylor expansion of the cosine one has $\Delta\phi^2\simeq
2(1-\<c\>)$. In Fig. \ref{f:asym} we plot $2(1-\<c\>)N$ of our optimal
estimation for $\theta=0$ versus $r$ for $N=16,18,20$, against the
Cramer-Rao bound $\frac{1}{r^2}$. From the comparison we see that our
estimation approaches the Cramer-Rao bound for large $N$. Notice that
according to recent studies of theoretical statistics \cite{gill},
there should exist a separable strategy (such as an adaptive scheme)
which is not necessarily covariant, nevertheless it would be able to
achieve the same Cramer-Rao bound asymptotically: such non covariant
schemes, e. g. homodyne-based estimation of the phase, will be the
subject of further studies.\par

\section{Conclusions}
In conclusion, we have presented the optimal measurement for phase
estimation on $N$ qubits all prepared in the same arbitrary mixed
state. The Uhlman fidelity saturates the Cramer-Rao bound for this
problem, confirming the optimality of the measurement. The optimal
estimation is achieved for equatorial qubits and generally the
fidelity is improving with purity. The specific form of the optimal
POVM in terms of the generalized Susskind-Glogower vector in Eqs.
(\ref{Susskind}) and (\ref{singlets}) suggests possible physical
implementations in terms of a generalized multipartite Bell
measurement.
\acknowledgements
\par We are grateful to R. Gill for pointing us the asymptotic behavior in terms of the Fisher information.
This work has been co-founded by the EC under the program ATESIT (Contract No.
IST-2000-29681) and the MIUR cofinanziamento 2003. P.P. acknowledges support from the INFM under
project PRA-2002-CLON. G.M.D. acknowledges partial support by the MURI program administered by
the U.S. Army Research Office under Grant No. DAAD19-00-1-0177.

\end{document}